\documentclass[12pt]{article}

\usepackage[dvips]{graphicx}
\usepackage{amsmath}
\usepackage{amsfonts}
\usepackage{placeins}
\usepackage{afterpage}
\usepackage{cite}
\usepackage{flafter}

\usepackage{dcolumn}

\usepackage[a4paper]{typearea} % AusgangspapiergrM-vM-_e
  \areaset[0cm]%                 % ZusM-dtzlicher Rand fuer die Bindung
          {17.0cm}{24cm}           % Textbreite und HM-vhe

\newcommand{\TABB}[5][AUSFUELLEN]{\begin{table}[#2]

     \protect#3

     \centerline{\parbox{16.5cm}{\caption[#1]{{\small #4}} \label{#5}}}

     \end{table} }

  \newcommand{\PSImagx}[2]{\includegraphics[width=#2]{psplots/#1}}

\newcommand{\BILD}[4]{\begin{figure}[#1]%

     #2

     \centerline{\parbox{16.5cm}{\caption[.]{{\small #3}} \label{#4}}}
     \end{figure} }

\newcommand{\Int}{\int\limits}

\newcommand{\ud}{\text{d}}
\newcommand{\ui}{\text{i}}

\newcommand{\R}{\mathbb{R}}
\newcommand{\Z}{\mathbb{Z}}

\begin{document}

\vspace*{-1cm}

ULM--TP/01--03

HPL--BRIMS--2001--10

April 2001

\vspace{1.5cm}

%%%%%%%%%%%%%
\newcommand{\titel}{Amplitude distribution of\\[1ex] 
                    eigenfunctions in mixed systems}
%%%%%%%%%%%%%

\normalsize

\vspace{0.5cm}

\begin{center}  \LARGE  \bf

      \titel
\end{center}

   \vspace{2.0cm}

\begin{center}

          {\large Arnd B\"acker%
\footnote[1]{E-mail address: {\tt a.backer@bristol.ac.uk}}$^{,2}$
          }
          {\large and Roman Schubert%
\footnote[3]{E-mail address: {\tt roman.schubert@physik.uni-ulm.de}}
           }

    \vspace{1.5cm}

    $^{1)}$ 
 School of Mathematics, University of Bristol\\
 University Walk, Bristol BS8 1TW, UK\\

    \vspace{2ex}

 $^{2)}$ BRIMS, Hewlett-Packard Laboratories\\
 Filton Road, Bristol BS12 6QZ, UK\\

    \vspace{2ex}

    $^{3)}$ Abteilung Theoretische Physik, Universit\"at Ulm,\\ 
    Albert-Einstein-Allee 11, D--89069 Ulm, Germany\\
    \vspace{4ex}

\end{center}

\renewcommand{\thefootnote}{\arabic{footnote}}
\setcounter{footnote}{0}

%%%%%%%%%%%%%%%%%%%%%%%%%%%%%%%%%%%%%%%%%%%%%%%%%%%%%%%%%%%%%%%%%%%%%%%%%%%%%

\newcommand{\erf}{\operatorname{erf}}
\newcommand{\BFq}{{\bf q}}
\newcommand{\BFx}{{\bf x}}
\newcommand{\BFp}{{\bf p}}
\newcommand{\BFk}{{\bf k}}
\newcommand{\BFe}{{\bf e}}
\newcommand{\noBFq}{q}
\newcommand{\limacon}{lima\c{c}on }
\newcommand{\omm}{s}
\newcommand{\BFn}{\boldsymbol n}
\newcommand{\E}{\operatorname{\mathbb{E}}}

\newcommand{\cP}{{\cal P}}
\newcommand{\cD}{{\cal D}}
\newcommand{\BFT}{{\bf T}}
\newcommand{\BFt}{{\bf t}}

\newcommand{\cL}{{\cal L}}

\newcommand{\VARPHI}{\phi}

\newcommand{\BFa}{{\bf a}}

\newcommand{\Vol}[1]{{\text{vol}(#1)}}
\newcommand{\psiRWM}{\psi_{\text{RRWM}}}
\newcommand{\psiRWMD}{\psi_{\text{RRWM},D}}
\newcommand{\PRWMDpsi}{P_{\text{RRWM},D}(\psi)}
\newcommand{\PRWMpsi}{P_{\text{RRWM}}(\psi)}

\newcommand{\EPS}{\varepsilon}
\newcommand{\var}{\sigma^2}

\vspace*{3cm}

\leftline{\bf Abstract:}

\noindent
We study the amplitude distribution of irregular eigenfunctions
in systems with mixed classical phase space.
For an appropriately restricted random wave model
a theoretical prediction for the  amplitude distribution  
is derived and good agreement with numerical computations 
for the family of \limacon billiards is found.
The natural extension of our result to more general systems, e.g.\
with a potential, is also discussed.

\newpage

%%%%%%%%%%%%%%%%%%%%%%%%%%%%%%%%%%%%%%%%%%%%%%%%%%%%%%%%%%%%%%%%%%%%%%%%%%%%%%%
\section{Introduction}
%%%%%%%%%%%%%%%%%%%%%%%%%%%%%%%%%%%%%%%%%%%%%%%%%%%%%%%%%%%%%%%%%%%%%%%%%%%%%%%

The semiclassical behavior of the eigenfunctions of a quantum 
mechanical system strongly depends on the ergodic properties of the 
underlying classical system. The semiclassical eigenfunction 
hypotheses \cite{Ber77b, Vor79} states that the Wigner function 
of a semiclassical
eigenstate is concentrated on a region in phase space 
explored by a typical trajectory of the classical system. 
In integrable systems the phase space is foliated into invariant tori, 
and the Wigner functions of the quantum mechanical eigenfunctions 
tend to delta functions on these tori in the 
semiclassical limit \cite{Ber77a}. 
On the other hand, in an ergodic system almost all trajectories 
cover the energy shell uniformly, and hence the Wigner functions 
of the eigenstates are expected to become a delta function on the energy 
shell. That this actually happens for an ergodic system for almost all 
eigenstates follows from the quantum ergodicity theorem 
see  \cite{Shn74,Zel87,CdV85},
and \cite{GerLei93,ZelZwo96} for billiards 
(the relation of the quantum ergodicity theorem 
to the semiclassical behaviour of Wigner functions is explicitly 
derived for Hamiltonian systems in \cite{BaeSchSti98}).
However, a generic system is neither integrable nor ergodic 
\cite{MarMey74}, but has a 
mixed phase space in which regular regions
(e.g.\ islands around stable periodic orbits)
and stochastic regions coexist.
Whether these numerically observed stochastic regions
are ergodic and of positive measure is an open question, see \cite{Str91} for
a review on the coexistence problem.
The eigenfunctions in mixed systems are expected
to be separated into regular and irregular eigenfunctions
according to an early conjecture by Percival \cite{Per73}
which has been numerically confirmed for several systems,
see e.g.\ \cite{BohTomUll90a,ProRob93b,LiRob95b,CarVerFen98}.
In addition, at finite energies
there is a small (semiclassically vanishing) fraction
of ``hierarchical states'' which are of intermediate nature,
and localize in regions bounded by cantori \cite{KetHufSteWe2000}.

Beside the localization properties of the Wigner function, also the 
local amplitude fluctuations of the eigenfunctions strongly depend  
on the classical system, as has been pointed 
out in \cite{Ber77b,Ber83}. The basic idea is that an eigenfunction can 
be represented locally as a superposition of de Broglie waves
with wavelength determined by the energy, and momenta distributed according
to the 
semiclassical limit of the Wigner function. In a chaotic system one 
therefore expects an isotropic distribution of the momenta. If one 
additionally assumes that the phases are randomly distributed, one obtains 
locally a Gaussian amplitude distribution of a typical eigenfunction in a 
quantum mechanical system with chaotic classical limit. 
For instance in a chaotic billiard a Gaussian amplitude distribution 
is expected, and this has been confirmed by 
several numerical studies, 
see e.g.~\cite{ShaGoe84,McDKau88,AurSte91,HejRac92,AurSte93,LiRob94}. 
Furthermore, predictions of the random wave model for the maxima of
chaotic eigenfunctions have been derived and numerically tested for
several systems in \cite{AurBaeSchTag99}.
In contrast, in an integrable system the localization of the 
Wigner function on the invariant tori enforces 
a more coherent superposition 
of the de Broglie waves, leading 
to a regular structure of the eigenfunction 
\cite{Ber77b}.

Our aim is to determine the amplitude distribution for irregular states 
in systems with a mixed classical dynamics. 
We assume  that the motion on a stochastic region 
$D$ in phase space 
is ergodic and that the statistical 
properties of eigenfunctions  
can be described by a random wave model restricted to $D$
 see the 
following section for a precise definition. 
The derivation shows that locally the fluctuations are Gaussian with a 
position dependent variance which is given by the classical
probability density on position space defined by the ergodic 
density on $D$. 
Thus the resulting amplitude distribution may be significantly
different from a Gaussian. In section \ref{sec:comparision}
we compare the theoretical prediction of the restricted random wave model
with numerical computations.

%%%%%%%%%%%%%%%%%%%%%%%%%%%%%%%%%%%%%%%%%%%%%%%%%%%%%%%%%%%%%%%%%%%%%%%%%%%%%%%
\section{Amplitude distribution for the restricted random wave model}
\label{sec:}
%%%%%%%%%%%%%%%%%%%%%%%%%%%%%%%%%%%%%%%%%%%%%%%%%%%%%%%%%%%%%%%%%%%%%%%%%%%%%%%

In this section  
we consider a restricted random wave model for 
two-dimensional Euclidean quantum billiards in order 
to describe the statistical properties of
irregular eigenfunctions in systems with a  
mixed classical phase space. 
The quantum mechanical system is defined by the 
Euclidean Laplacian on a compact domain $\Omega \subset \R^2$ 
with suitable boundary conditions 
on the boundary $\partial \Omega$. (Usually one chooses Dirichlet 
conditions.) The quantum mechanical 
eigenvalue problem is given by 
\begin{equation}
-\Delta \psi_n(\BFq)=E_n \psi_n(\BFq)\,\, ,\qquad\text{with}\quad
\psi_n(\BFq)=0 \quad  \text{for $\BFq \in{\partial \Omega}$}\,\, , 
\end{equation}
and we are interested in the behavior of the eigenfunctions $\psi_n$ in the 
semiclassical limit $E_n\to\infty$. 

The corresponding classical system is given by a free particle moving 
along straight lines inside the billiard, making elastic reflections
on the billiard boundary $\partial \Omega$. 
The phase space is $T^*\Omega=\R^2\times \Omega$, 
and the Hamiltonian is $H(\BFp,\BFq)=|\BFp|^2$. Since the 
Hamiltonian is scaling we can restrict our attention to the 
equi-energyshell with energy $E=1$, 
\begin{equation}
S^*\Omega:=\{ (\BFp,\BFq) \in  \R^2\times \Omega\;\, ; \;\, |\BFp|=1\}\,\, . 
\end{equation} 
Introducing polar coordinates $(r,\VARPHI)$ for the momentum $\BFp$, 
we can parametrize $S^*\Omega$ by $(\VARPHI, \BFq)\in [0,2\pi)\times \Omega$ 
where $\VARPHI$ is the direction of the momentum. In these coordinates the 
Liouville measure on $S^*\Omega$ is given by
\begin{equation}\label{eq:def-Liouville-measure}
\ud \mu =\ud \VARPHI \,\ud^2 q\,\, ,
\end{equation}  
which is invariant under the Hamiltonian flow on $S^*\Omega$.

Now let $D\subset S^*\Omega$ be an open domain which is invariant under 
the classical flow, and on which the flow is chaotic. 
The existence 
of such a domain where the flow is, for instance, ergodic, is an open problem.
But numerically one observes invariant domains on which the flow is 
at least irregular in the sense that most orbits are unstable, and 
regular islands inside this domain are very small. 
The uncertainty principle implies a finite quantum mechanical resolution 
of phase space quantities at finite energies.
Therefore at finite energies 
the small islands of such an irregular domain 
are not resolved by the quantum system.

So we expect, in the spirit of \cite{Ber77b}, that 
the statistical properties of irregular eigenfunctions associated with $D$ 
can be described by those of a superposition of plane waves 
with wave vectors of the same lengths and directions distributed uniformly on 
$D$. If we furthermore assume random phases we arrive for real valued 
functions at the following restricted random wave model, 
which is a superposition of plane waves of the form 
\begin{equation}\label{eq:restricted_random_wave}
  \psiRWMD (\BFq) 
   = \sqrt{\frac{4\pi}{\Vol{D} N}}
          \sum_{n=1}^{N}  \chi_D(\widehat{\BFk}_n, \BFq)
               \cos(\BFk_n \BFq + \EPS_n) \;\;.
\end{equation}
Here $\chi_D(\cdot)$ is the characteristic function of $D$, 
the phases $\EPS_n$ are independent random 
variables equidistributed on $[0,2\pi]$, and the momenta 
$\BFk_n\in\R^2$ are independent random variables which are 
equidistributed on the circle
of radius $\sqrt{E}$. So the characteristic 
function $\chi_D(\cdot)$ ensures the localization on $D$.
Furthermore, it is natural to take $N\sim\sqrt{E}$,
the scaling of the number of line
segments of a typical Heisenberg--length orbit.
By $\Vol{D}$ the volume of $D$ measured with the Liouville 
measure \eqref{eq:def-Liouville-measure} is denoted.  
With this choice of normalization the expectation value 
of the norm $||\psiRWMD||$ is one. 

Let us first consider the value distribution  $P_{\BFq}(\psi)$ 
of $\psiRWMD(\BFq)$
at a given point $\BFq\in\Omega$.
Our restricted random wave model \eqref{eq:restricted_random_wave}  
is a sum of  identical independent random variables 
which have zero mean and whose variance is given by 
\begin{equation}\label{eq:local_variance}
\begin{split}
\var(\BFq)&=\E\bigg(\frac{4\pi}{\Vol{D}}\big( \chi_D(\widehat{\BFk}_n, \BFq)\cos(\BFk_n \BFq + \EPS_n)\big)^2\bigg)\\
&=\frac{1}{\Vol D}\, 
        \Int_{0}^{2\pi} \chi_D(\BFe(\VARPHI),\BFq) \; \ud\VARPHI \;\;,
\end{split}
\end{equation}
where $\BFe(\VARPHI):=(\cos(\VARPHI),\sin(\VARPHI))$ denotes the unit vector 
in $\VARPHI$-direction.  So by the central limit theorem we  obtain 
for $E\to\infty$, i.e.\ $N\to\infty$, a Gaussian   
distribution of $\psiRWMD(\BFq)$ at $\BFq$, 
\begin{align}  \label{eq:value-distrib-q-restricted-random-wave}
  P_{\BFq}(\psi)  \longrightarrow 
        \sqrt{\frac{1}{2\pi\var(\BFq)}} \exp\left(
          -\frac{\psi^2}{2\var(\BFq)}\right) \;\; ,
\end{align} 
with variance given by \eqref{eq:local_variance}. 
If the classical dynamics on $D$ is ergodic, then the  variance 
$\var(\BFq)$ is exactly 
the probability density of finding the particle 
in the point $\BFq\in \Omega$ if 
it moves on a generic trajectory in $D$. So $\var(\BFq)$ is the classical 
probability density in position space.

By integrating eq.~\eqref{eq:value-distrib-q-restricted-random-wave}
over $\Omega$ we obtain the complete amplitude 
distribution as a mean over a family of Gaussians with variances given 
by \eqref{eq:local_variance},
\begin{align} \label{eq:amp-distrib-mixed}
  P_{\text{RRWM},D} (\psi) & = \frac{1}{\Vol{\Omega}} 
                 \Int_\Omega P_{\BFq}(\psi)  \; \ud^2 \noBFq \\
         &= \frac{1}{\Vol{\Omega}} 
                 \Int_\Omega \sqrt{\frac{1}{2\pi\var(\BFq)}}
           \exp\left(-\frac{1}{2\var(\BFq)}\psi^2 \right)   
        \; \ud^2 \noBFq \,\, .\label{eq:amp-distrib-mixed-}
\end{align} 
So the amplitude distribution is completely determined by the classical 
probabality density \eqref{eq:local_variance}, and it will be typically 
non Gaussian if $\var(\BFq)$ is not constant. 

The moments of the distribution \eqref{eq:amp-distrib-mixed-}
can be computed 
directly and turn out to be proportional to 
the moments of the classical probability density $\var(\BFq)$, 
\begin{equation} \label{eq:moments-via-sigma-q}
\int \psi^{2k} P_{\text{RRWM},D} (\psi)\,\, \ud \psi =
\rho_{2k}\, \frac{1}{\Vol\Omega}\Int_{\Omega}[\var(\BFq)]^k\,\, \ud q  \;\;,
\end{equation}
where the factor $\rho_{2k}=\frac{(2k)!}{k!2^k}$ 
denotes the $2k$'th moment of a Gaussian. The odd 
moments are of course zero. Note that the second moment is always 
$1/\Vol\Omega$,  due to the normalization of $\psi$.

If the system is ergodic one has $\var(\BFq)=\frac{1}{\Vol{\Omega}}$
and we get the classical result that $\PRWMDpsi$ is Gaussian with
variance $\var=\frac{1}{\Vol{\Omega}}$.
However, if $\var(\BFq)$ depends on $\BFq$
then the corresponding distribution can show deviations
from the Gaussian distribution.
In particular, if $\var(\BFq)=0$ for some region $\Omega'\subset\Omega$,
we get a contribution $\frac{\Vol{\Omega'}}{\Vol{\Omega}} \delta(\psi)$
to the corresponding distribution of $\PRWMDpsi$
as the integrand in \eqref{eq:amp-distrib-mixed}
tends to a $\delta$ distribution as $\var(\BFq)\to 0$.

Finally, we would like to point out that the main ingredient in the 
formula \eqref{eq:amp-distrib-mixed} is the assumption that the 
local amplitude distribution of an irregular eigenfunction 
around a point $\BFq$ in position space 
is Gaussian, with a variance given by the classical probability density 
in position space $\var(\BFq)$, 
defined by the projection of the invariant measure 
on $D$ to position space. Clearly this assumption is not restricted to 
billiards, but is expected to be true for 
arbitrary quantum mechanical systems for which the underlying classical 
system contains chaotic components in phase space. So the formula 
\eqref{eq:amp-distrib-mixed} is expected to be valid in far more general 
situations, with $\var(\BFq)$ denoting the classical probability density 
defined by the ergodic measure on the chaotic component.

%%%%%%%%%%%%%%%%%%%%%%%%%%%%%%%%%%%%%%%%%%%%%%%%%%%%%%%%%%%%%%%%%%%%%%%%%%%%%%%
\section{Comparison with irregular eigenfunctions} \label{sec:comparision}
%%%%%%%%%%%%%%%%%%%%%%%%%%%%%%%%%%%%%%%%%%%%%%%%%%%%%%%%%%%%%%%%%%%%%%%%%%%%%%%

We now compare the predictions of the restricted random wave model
with the 
results for some numerically computed eigenfunctions. 
As systems to study the amplitude distribution of
irregular states in mixed systems we have chosen the
family of \limacon billiards introduced by Robnik \cite{Rob83,Rob84}
with boundary given in polar coordinates by 
$\rho(\varphi)=1+\varepsilon \cos(\varphi)$, $\varphi\in[-\pi,\pi]$,
with $\varepsilon \in [0,1]$  
being the system parameter. We consider the case
$\varepsilon=0.3$, for which the billiard
has a phase space of mixed type \cite{Rob83},
see fig.~\ref{fig:poincare}.
In \cite{LiRob95} examples of eigenstates far in the 
semiclassical regime have been studied in this system, and
in particular the amplitude distribution has been 
studied numerically, but no analytical predictions have been made.

\BILD{b}
     {
      \begin{center}
         \PSImagx{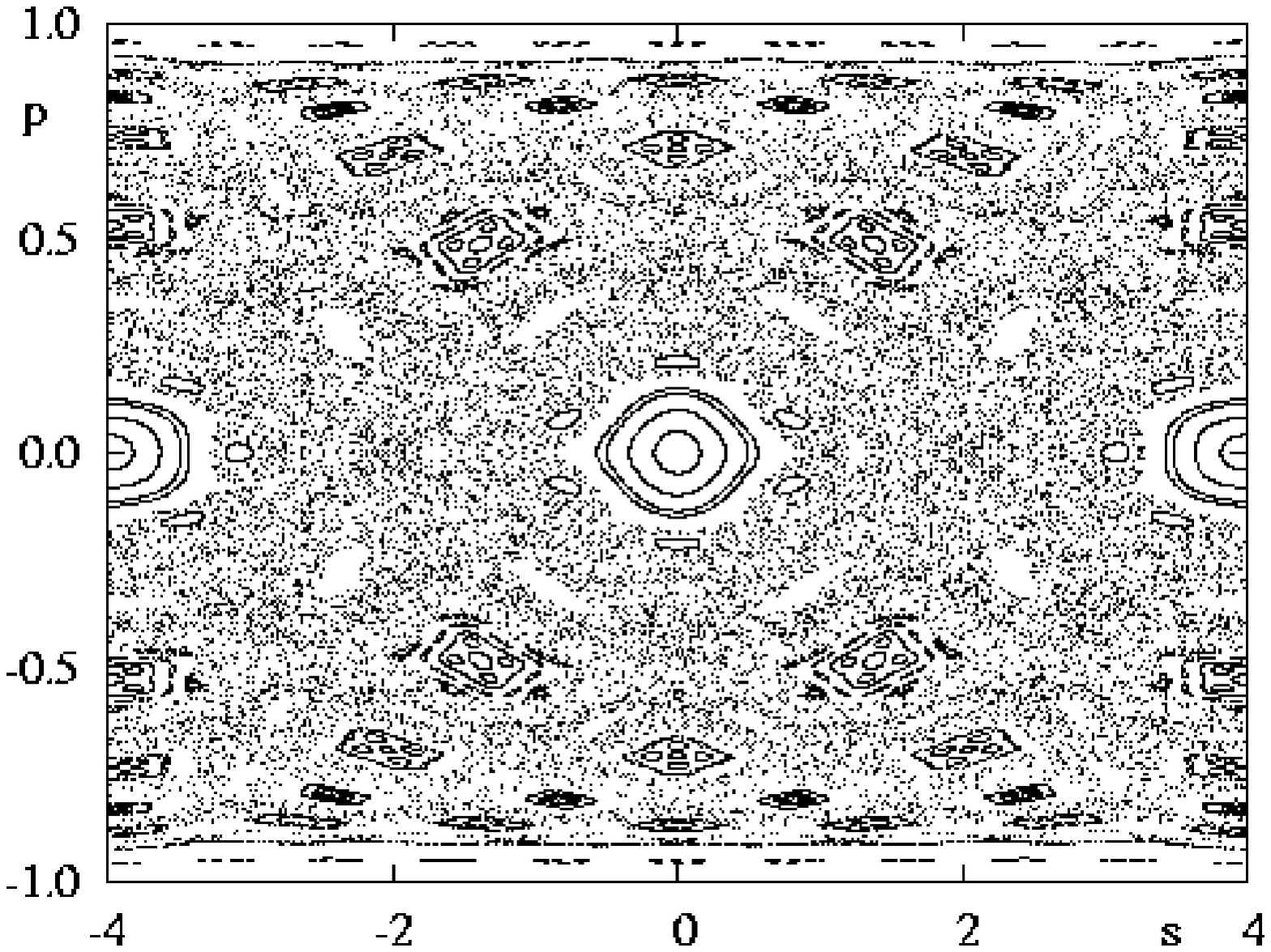}{14cm}
    \end{center}
     }
     {Plot of several stable and irregular orbits in the
      Poincar\'e section $\cP$
      of the \limacon billiard for $\varepsilon=0.3$.
      Here $\cP$ is parametrized by the (rescaled) arclength coordinate
      $s\in [-4,4]$ 
      along the billiard boundary
      and $p\in[-1,1]$ which is the projection 
      of the unit velocity vector on the tangent in the
      point $s$ after the reflection.}
     {fig:poincare}

First we have to determine the classical position space probability density 
$\var(\BFq)$ of the ergodic measure on the invariant domain $D$. 
The normalized ergodic measure on $D$ is given by 
\begin{equation}\notag
\ud \mu_{D}(\VARPHI,\BFq) =
       \frac{1}{\Vol D}\chi_D(\BFe(\VARPHI), \BFq)\, 
          \ud\VARPHI \,\ud^2\noBFq\,\, ,
\end{equation}
so we can express the variance $\var(\BFq)$ as a mean value 
\begin{equation}
\var(\BFq)=\Int_{S^* \Omega} 
       \delta (\BFq-\BFq') \,\, \ud \mu_{D}(\VARPHI ',\BFq') \;\;.
\end{equation}
As the motion on $D$ is assumed to be ergodic,
in order to determine $\var(\BFq)$ 
we could replace the integral over $S^* \Omega$ by
a time average over a typical trajectory of $D$ 
and the $\delta$ function
by a smoothened $\delta$ function, e.g.\ a narrow Gaussian.
However, as we will see below, the eigenfunctions turn out 
to be  not concentrated on the whole chaotic component, but rather 
on a subset which is almost invariant in the sense that it is bounded
by partial barriers in phase space. Since at finite 
energies quantum mechanics has only a finite resolution in phase space, 
these partial barriers appear like real barriers.
But since  any 
classical trajectory will pass such a barrier after a certain time, 
the time average is not suitable for the determination of $\var(\BFq)$
in such a situation.

For a more direct approach to determine 
$\var(\BFq)$ we use the Poincar\'e section 
$\cP = \{ (s,p) \;;\; s\in[-4,4],\; p\in[-1,1]  \}$,
which is parametrized by the (rescaled) arclength coordinate $s$ (corresponding
to $\varphi\in[-\pi,\pi]$) along the 
boundary $\partial \Omega$ and
the projection $p$ of the unit velocity vector on the tangent in the
point $s$ after the reflection.
Let $\cD \subset \cP$ be the projection of the region $D$ in the energy shell 
$S^*\Omega:=\{ (\BFp,\BFq)\in \R^2\times \Omega\;\, ; \;\,||\BFp||=1\}$
to the Poincar\'e section. This projection is defined as follows: 
to a point $(\BFe(\VARPHI) ,\BFq)\in D$ we can associate the trajectory 
which passes through $\BFq$ in direction $\BFe(\VARPHI)$, then 
$s(\VARPHI ,\BFq)$ is defined as the first intersection with the boundary 
$\partial \Omega$ when traversing the trajectory backwards from $\BFq$, 
and $p(\VARPHI ,\BFq):=\BFe(\VARPHI)\BFT(s(\VARPHI ,\BFq))$
which is the projection of the unit velocity vector $\BFe(\VARPHI)$
to the unit tangent vector $\BFT(s(\VARPHI ,\BFq))$ 
to $\partial \Omega$ at $s(\VARPHI ,\BFq)$.

For a given point $\BFq$ we therefore get a curve parameterized by $\VARPHI$ 
\begin{equation} \label{eq:p-phi-curve}
(p(\VARPHI ,\BFq), s(\VARPHI ,\BFq))\in \cP\,\, .
\end{equation}
Since $\chi_D(\BFe(\VARPHI),\BFq)
       =\chi_{\cD}((p(\VARPHI ,\BFq), s(\VARPHI ,\BFq))$, we get 
\begin{equation}
\var(\BFq)=\frac{1}{\Vol D}\, 
        \Int_{0}^{2\pi} \chi_{\cD}((p(\VARPHI ,\BFq), 
         s(\VARPHI ,\BFq)) \; \ud\VARPHI \;\;, 
\end{equation}
and therefore we
have to determine the fraction of the angular interval(s) for which the curve 
\eqref{eq:p-phi-curve} is in $\cD$.  
That is, one has to determine the angles
$\VARPHI_i^{\text{entry}}(\BFq)$ and $\VARPHI_i^{\text{exit}}(\BFq)$ where
the curve  \eqref{eq:p-phi-curve} enters or leaves the region $\cD$,
i.e.\ the intersection points of \eqref{eq:p-phi-curve}  with
the boundary of $\cD$.
In terms of these angles we obtain
\begin{equation} \label{eq:sigma-q-via-splines}
  \var(\BFq) =  \frac{1}{\Vol D}
          \sum_i  \VARPHI_i^{\text{exit}}(\BFq) 
                      - \VARPHI_i^{\text{entry}}(\BFq) \;\;,
\end{equation}
which is proportional to the fraction of directions in the ergodic component
visible from the point $\BFq$.

With this classical probability density $\var(\BFq)$
one can compute the corresponding amplitude distribution
via eq.~\ref{eq:amp-distrib-mixed-}.
If $\var(\BFq)=0$ for some region, then the local amplitude distribution 
\eqref{eq:value-distrib-q-restricted-random-wave} becomes a delta function, 
and it is necessary to 
consider for a concrete comparison a binned distribution,
\begin{align}
  P_{\text{binned}}(\psi,\Delta\psi) &:= \frac{1}{\Delta \psi}
       \Int_{\psi-\Delta\psi/2}^{\psi+\Delta \psi/2} P(\psi') \; \ud\psi' \\
      &= 
   \frac{1}{2|\Omega|}
            \Int_{\Omega}  
            \left[ \erf\left( \frac{\psi+\Delta\psi/2}
                                  {\sqrt{2\var(\BFq)}} \right) 
                  -
                   \erf\left( \frac{\psi-\Delta\psi/2}
                                  {\sqrt{2\var(\BFq)}} \right) 
            \right]  \;\ud^2\noBFq \;\;.
\end{align}

We now use a Husimi Poincar\'e section representation of
the eigenstate (see e.g.\ \cite{TuaVor95,SimVerSar97}) 
to determine the boundary of the relevant
component $\cD$ by a spline approximation.
The Poincar\'e Husimi representation 
of an eigenfunction $\psi_n$ in a billiard
is defined by projecting the normal derivative $u_n(s)$ of 
an eigenfunction $\psi_n(\BFq)$ 
at the boundary onto a coherent state 
on the boundary. The coherent states,
semiclassically centered in $(s,p)\in \cP$,
are defined as  
\begin{equation} \label{eq:coh-state-method0}
  c_{(s,p),k}(s') := 
      \left( \frac{k}{\sigma \pi}\right)^{1/4}
         \sum_{m=-\infty}^{\infty}
        \exp\left(\ui p k \left(s'-m L-s\right)
           \right)
        \exp \left( {- \frac{k}{2 \sigma}} 
               \left(s'-m L-s\right)^{2} \right) \;\;,
\end{equation}
where 
$s'\in [-4,4]$, $\sigma>0$ and $L=8$ is the total (rescaled) 
length of the boundary.
This definition is just a periodized version of 
the standard coherent states. 
The Poincar\'e Husimi function of a state $\psi_n$ with normal derivative 
$u_n(s)$ is then defined as 
\begin{equation} 
 H_n(s,p) =  
       \frac{k_n}{2\pi} 
      \frac{1}
           { \int_{-4}^{4} |u_n(s)|^2 \; \ud s } \left| \;
                  \Int_{-4}^4 c_{(s,p),k_n}^*(s') \; u_n(s')  \; \ud s' 
            \right|^2 \;\;,
\end{equation}
with $k_n=\sqrt{E_n}$; the prefactor ensures the normalization 
$\iint H_n(s,p) \,\, \ud p\,\ud s =1$. 

\BILD{t}
     {
      \begin{center}

        \vspace*{-1cm}
        \PSImagx{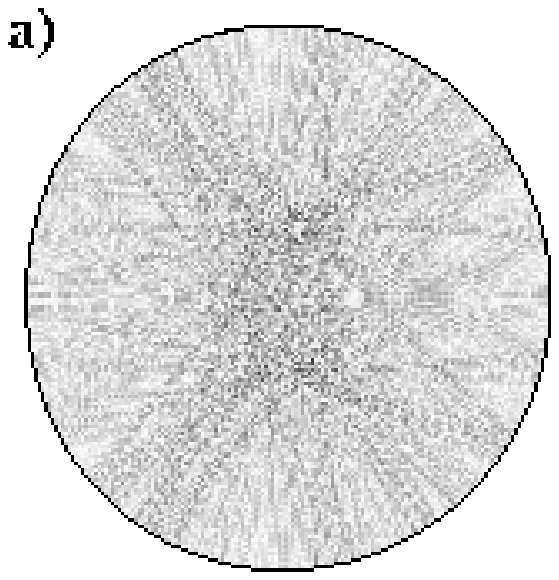}{5cm}

        \PSImagx{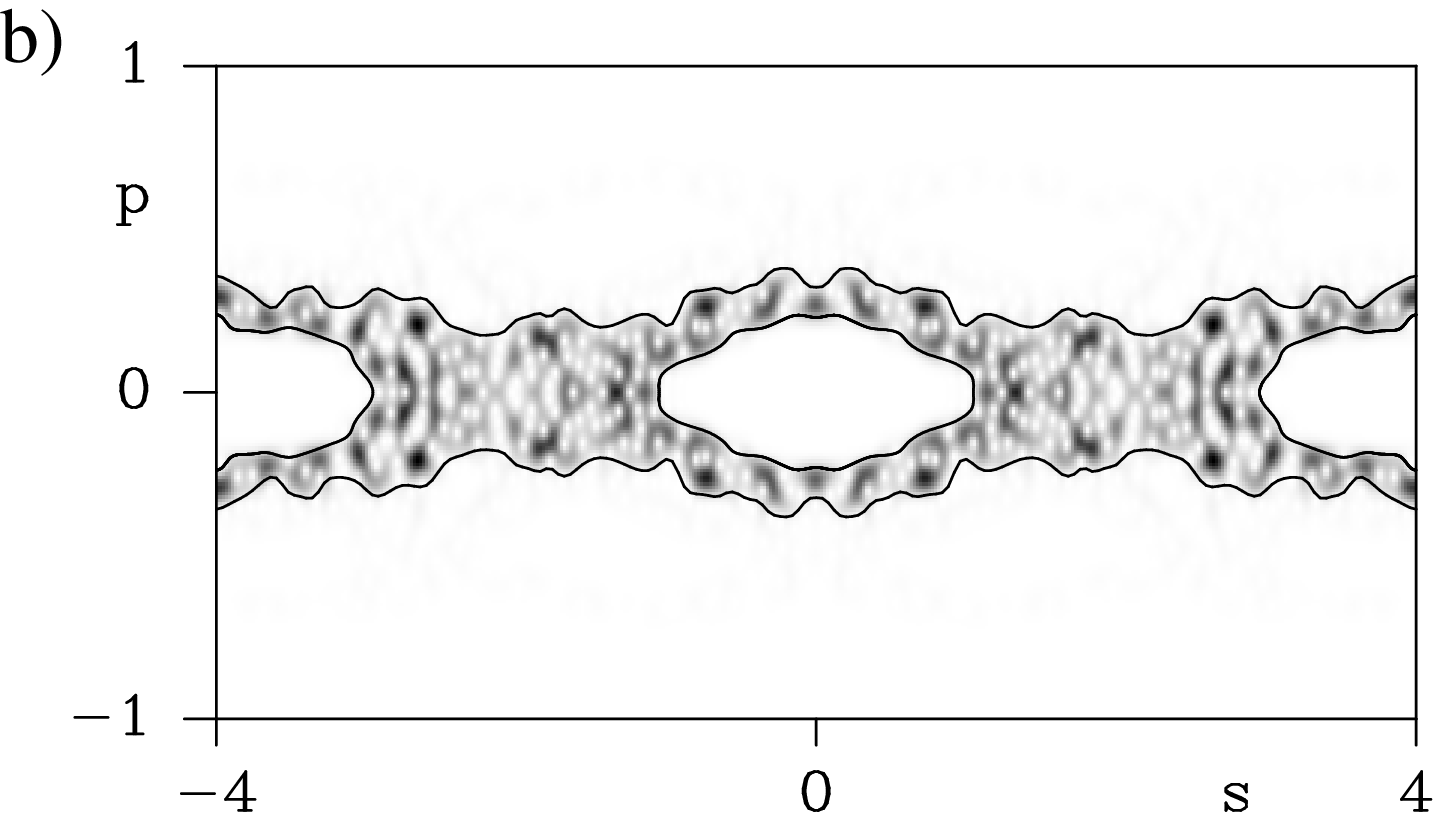}{10cm}
         \PSImagx{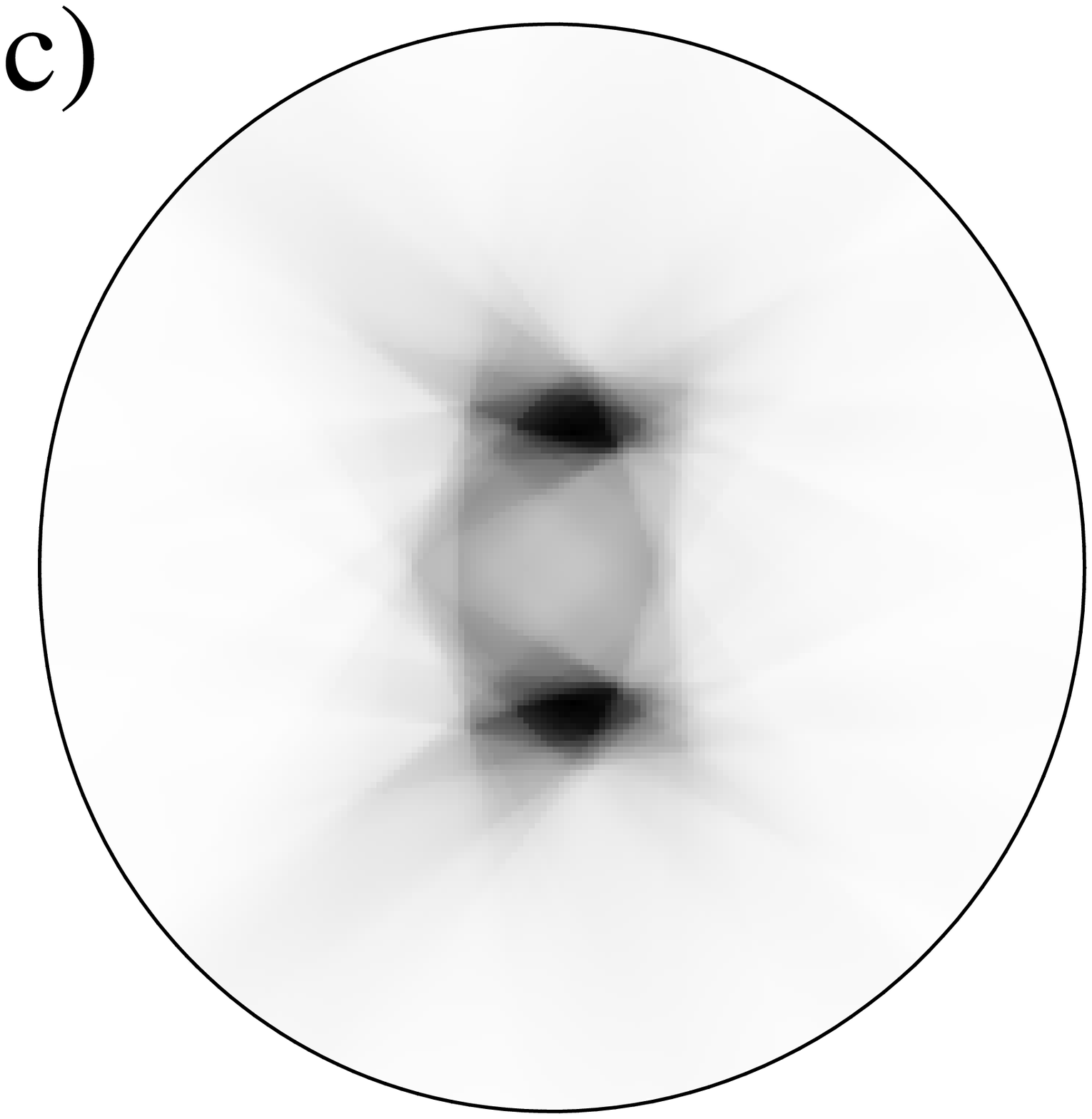}{5cm}

         \PSImagx{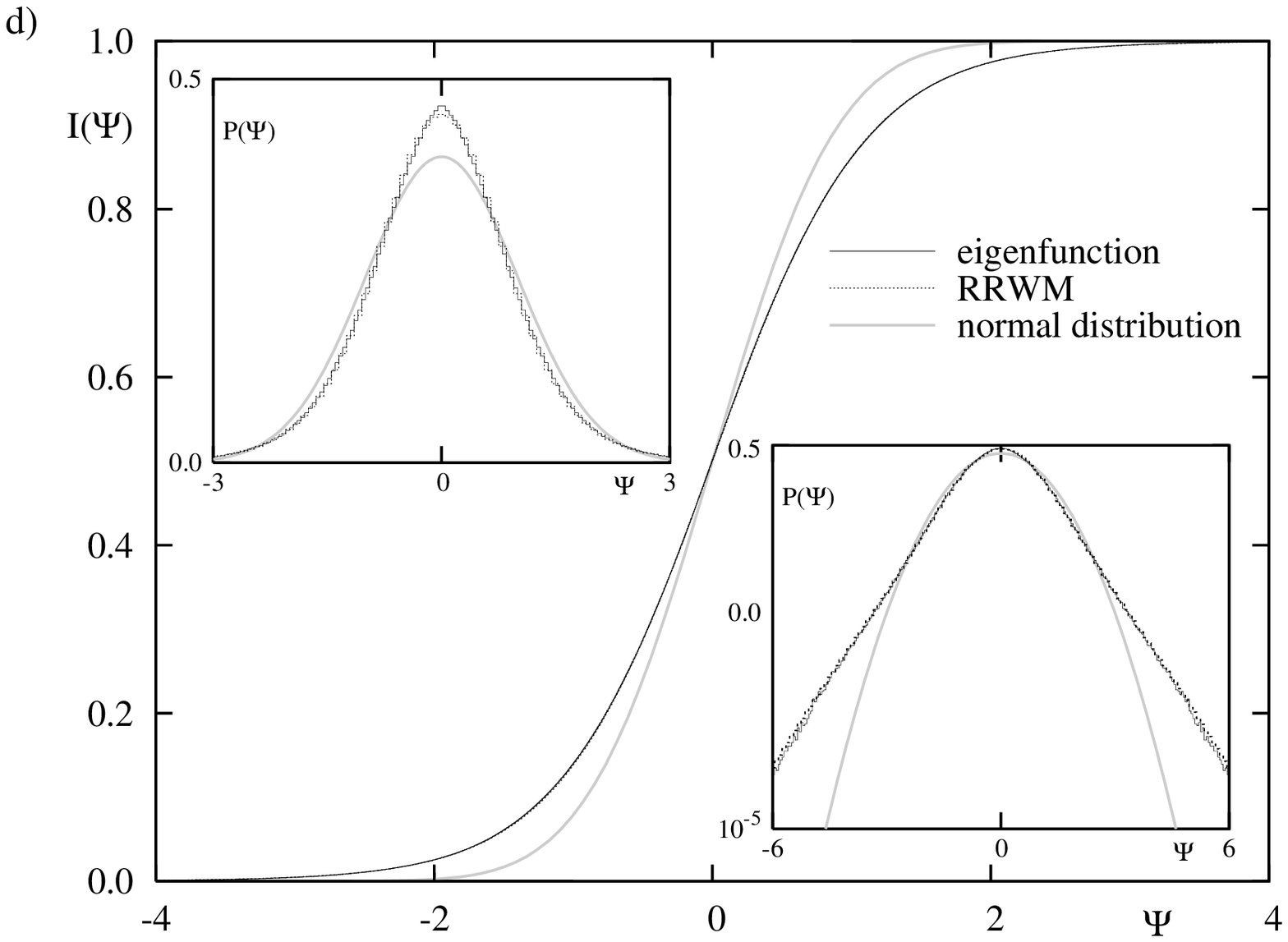}{16cm}

        \vspace*{-0.75cm}
     \end{center}
     }
     {In a) a high lying eigenfunction ($E=1002754.70\ldots$, 
      approximately the 130568$^{\text{th}}$ state) in the \limacon
      is shown as grey scale plot (black corresponding to high intensity).
      In b) the corresponding Husimi function on the Poincar\'e section
      is shown together with the boundary (full curves)
      of the region on which the eigenfunction is concentrated.
      In c) a density plot of $\var(\BFq)$, computed via 
       eq.~\eqref{eq:sigma-q-via-splines}, is shown.
      In d) the cumulative amplitude distribution
      of the eigenfunction is compared with the prediction
      of the restricted random wave model (RRWM); 
      on this scale no differences are visible.
      The left inset shows $P(\psi)$ and for the right inset
      a logarithmic vertical scale is used to emphasize the tails of 
      the distribution.
      For comparison the normal distribution is shown as grey curve.
}
     {fig:example1}

\BILD{t}
     {
      \begin{center}
        \vspace*{-1cm}

        \PSImagx{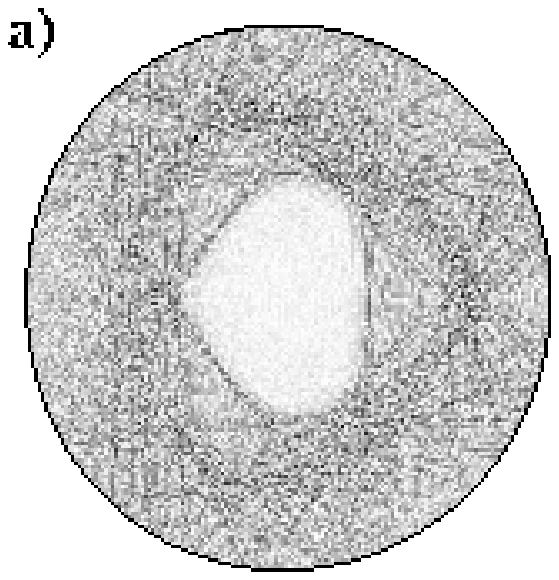}{5cm}

        \PSImagx{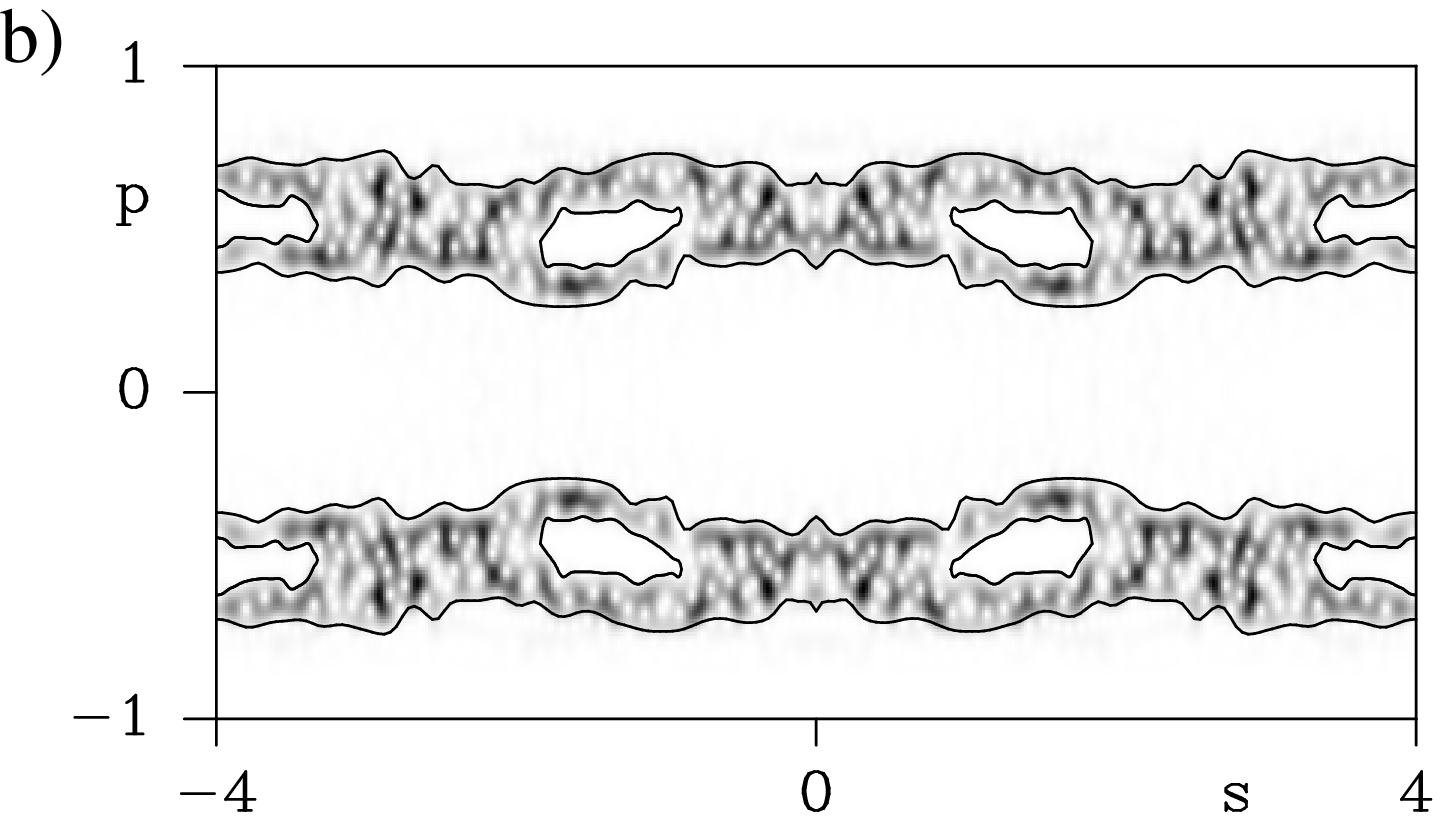}{10cm}
         \PSImagx{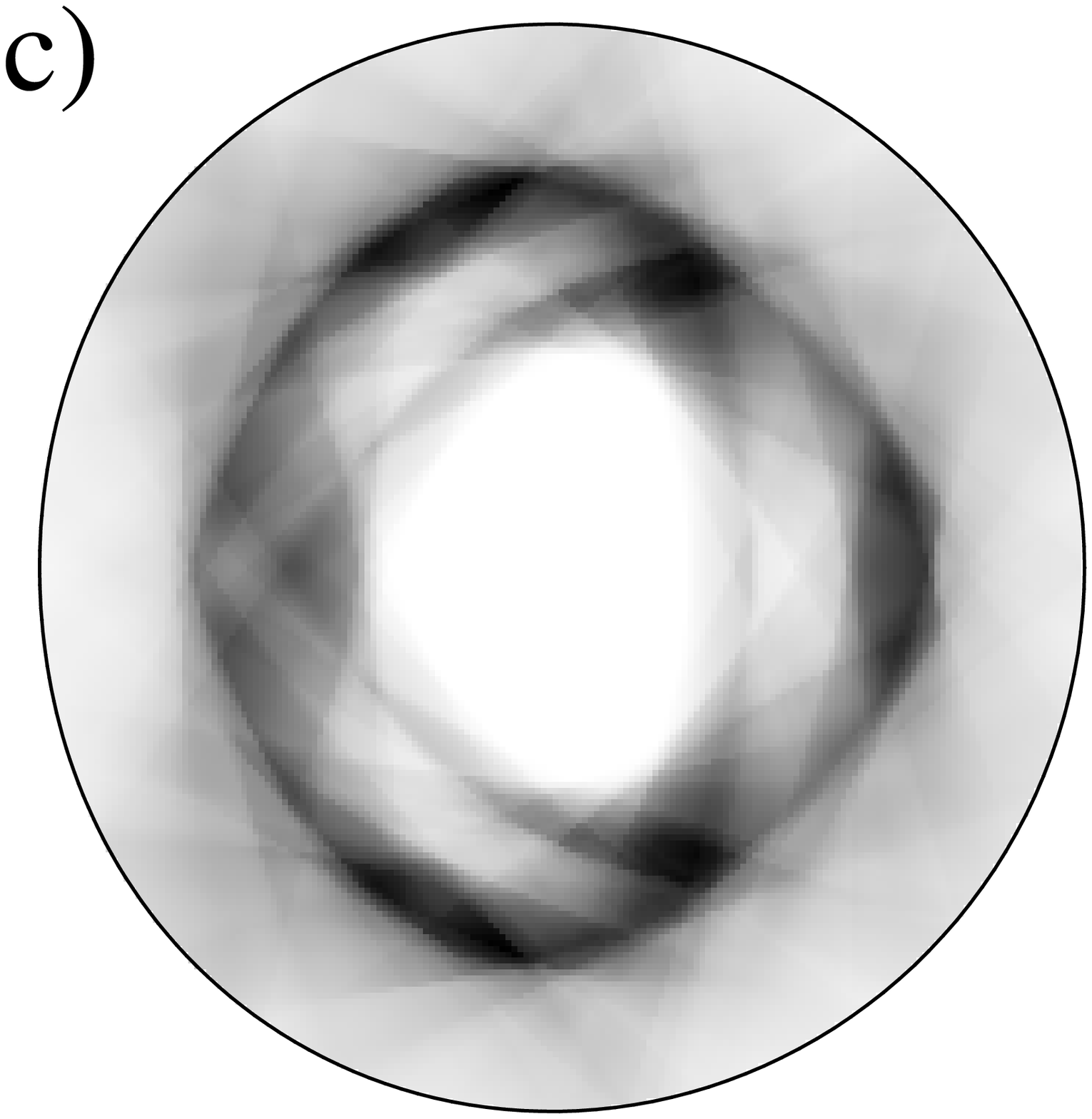}{5cm}

         \PSImagx{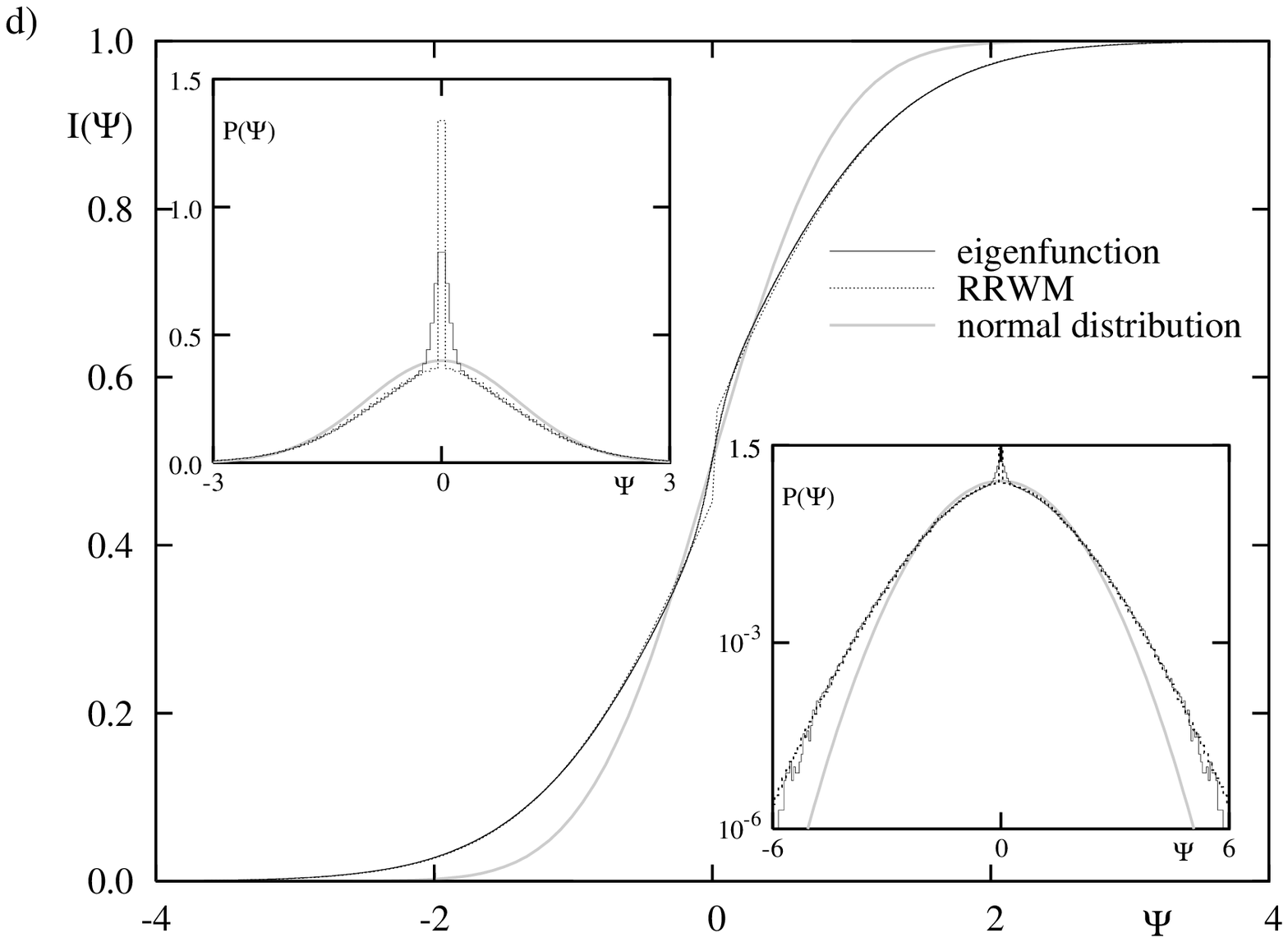}{16cm}

        \vspace*{-0.75cm}
     \end{center}
     }
     {The same plots as in the previous figure are shown
      for another high lying eigenfunction ($E=1003030.75\ldots$, 
      approximately the 130607$^{\text{th}}$ state).
      In this case there is a deviation of the amplitude distribution
      of the eigenfunction from the prediction of the restricted
      random wave model around $\psi=0$. This is because
      $\var(\BFq)=0$ in the central region, whereas the eigenfunction
      does not vanish there (see the text for further discussion).
      For the tails of the distribution the agreement
      of the two distributions is again very good.
    }
     {fig:example2}

\newcolumntype{d}[1]{D{.}{.}{#1}}

\setlength{\extrarowheight}{3pt}

\newcommand{\TextA}[1]{\multicolumn{1}{c|}{#1}}
\newcommand{\TextB}[1]{\multicolumn{1}{c||}{#1}}
\newcommand{\TextC}[1]{\multicolumn{1}{c||}{#1}}

\TABB{H}
     {

\begin{center}
     \begin{tabular}{c||d{-1}|d{-1}||d{-1}|d{-1}||c}
 & \multicolumn{2}{c||}{Example 1, fig.~\ref{fig:example1}} & \multicolumn{2}{c||}{Example 2, fig.~\ref{fig:example2}} \\
\raisebox{2ex}{moment} 
 & \TextA{eigenfunction} & \TextB{RRWM} & \TextA{eigenfunction} & \TextC{RRWM} & \raisebox{2ex}{normal distribution}\\

\hline
 \phantom{1}4 &     4.39 &    4.46   &    3.85 &    3.75                & \phantom{94}3 \\
 \phantom{1}6 &    45.1  &   47.6    &   26.9  &   25.8                 & \phantom{9}15 \\
 \phantom{1}8 &   819    &  899      &  269    &  269                   &           105 \\
           10 & \;\; 2199    & \;\;2501      & \;\;3774    & \;\;3841   &           945 \\
\end{tabular}
\end{center}
    }
     {Comparison of the even moments for the distributions of the 
      eigenfunction  and
      the restricted random wave model (RRWM) 
      eq.~\eqref{eq:moments-via-sigma-q}. The last column lists 
      for comparison 
      the moments of the normal distribution.}
     {tab:moments}

\BILD{b}
     {
      \begin{center}
        \PSImagx{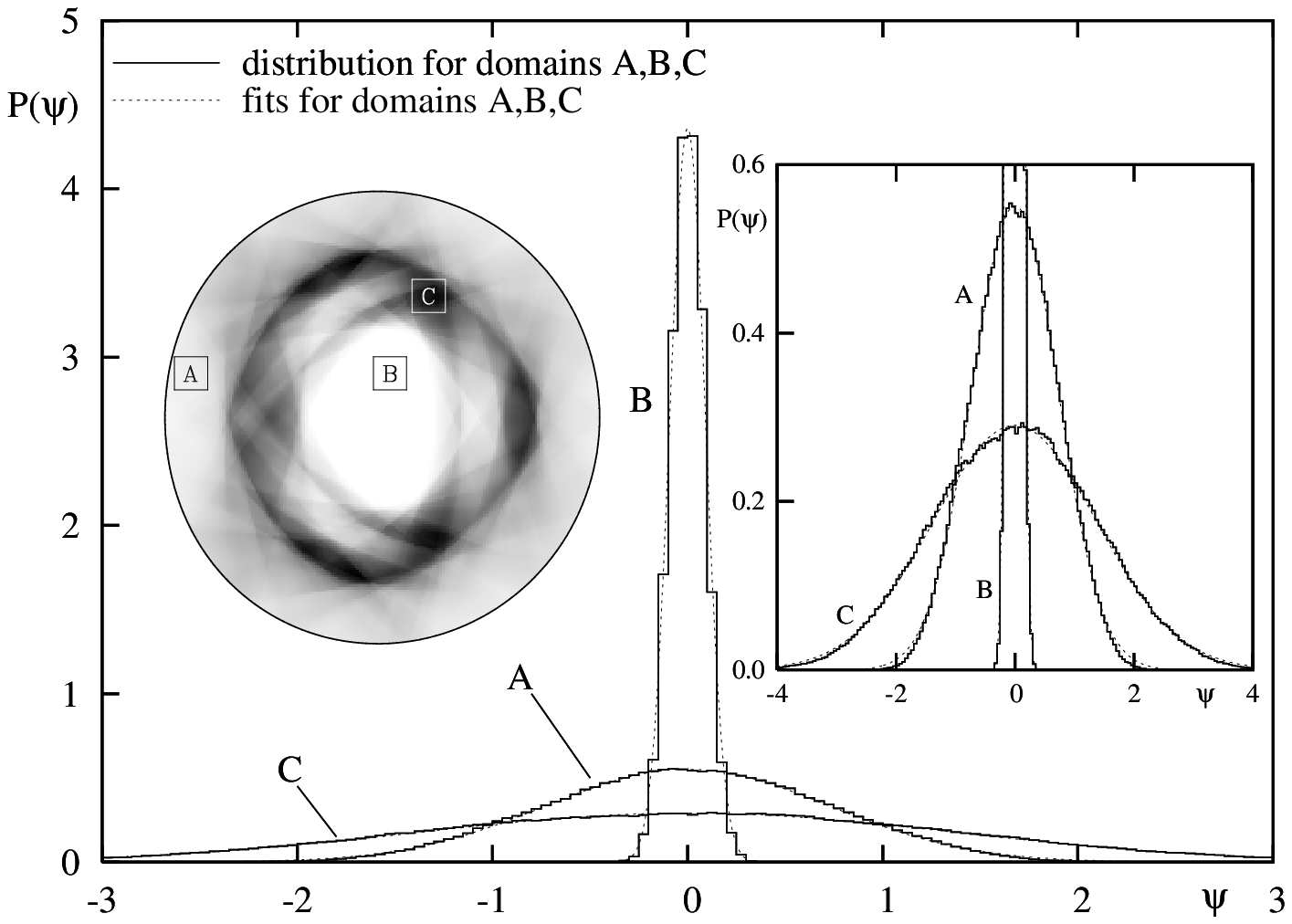}{17cm}
     \end{center}
     }
     {For the three domains indicated in the inset 
      the local amplitude distribution is shown (for the same state
      as in fig.~\ref{fig:example2}).
      We observe very good agreement with the expected Gaussian
      behaviour (shown as dotted curves) with position dependent variance.
      Notice that the non-zero width of the distribution for the region B
      corresponds to the widening of the $\delta$--contribution, 
      see fig.~\ref{fig:example2}.}
      {fig:local-fluctuations}

\afterpage{\clearpage}

A first example is shown in fig.~\ref{fig:example1}.
In a) a high lying eigenfunction 
($E=1002754.70\ldots$, approximately the 130568$^{\text{th}}$ state
of odd symmetry)
in the \limacon billiard with $\varepsilon=0.3$
is shown as density plot (black corresponding to high intensity of $|\psi|^2$).
In b) the corresponding Husimi representation on the Poincar\'e section
is shown.
The boundary of the irregular region $\cD$ is described
by a cubic spline which is shown as full curve.
With these boundary curves 
we can use \eqref{eq:sigma-q-via-splines} to compute
$\var(\BFq)$, which is shown in fig.~\ref{fig:example1}c).
Finally, in fig.~\ref{fig:example1}d) the comparison
of the amplitude distribution of $\psi$
with the prediction of the restricted random wave model is given.
Clearly, $P(\psi)$ is non--Gaussian,
and the agreement is very good.
Table \ref{tab:moments} lists the first moments 
and also very good agreement of the results using
\eqref{eq:moments-via-sigma-q} and the moments of $\psi$ is found.
Both the resulting amplitude distribution 
$P_{\text{RRWM},D}$ and the moments turn out to be quite robust
with respect to small changes of the selection of $\cD$.
Notice that we have rescaled $\var(\BFq)$
such that the variance of the distributions is one.

Another example is shown in fig.~\ref{fig:example2}. 
The eigenfunction ($E=1003030.75\ldots$, approximately the 
130607$^{\text{th}}$ state 
of odd symmetry) plotted in  a)
has a quite large region in the center where it is almost vanishing.
So from this alone the amplitude distribution is expected
to show a very clear deviation from the normal distribution.
Using the same procedure as in the previous case we
determine $\cD$, compute $\var(\BFq)$ and 
then $\PRWMpsi$.
The comparison of the prediction with $P(\psi)$
is shown in fig.~\ref{fig:example2}d).
The strongest deviation occurs for $\psi\approx 0$.
The peak of $\PRWMpsi$ at $\psi=0$ is due to the
fact that $\var(\BFq)=0$ for
the region in the center of the billiard.
The eigenfunction, however, is not exactly zero, but shows a
decay into that region and thus still fluctuates there.
This causes a broadening of the $\delta$--contribution,
which is clearly visible in the plot of $P(\psi)$ in 
fig.~\ref{fig:example2}d).
For $|\psi|>0.25$ this region is not relevant anymore,
and the agreement of $P(\psi)$ and $\PRWMpsi$ is
very good.
In the right inset of fig.~\ref{fig:example2}d) the distribution
is shown with a logarithmic vertical scale 
to illustrate the agreement of the distributions even in the tails.

The moments, computed via eq.~\eqref{eq:moments-via-sigma-q},
are listed in table \ref{tab:moments}.
The agreement of the moments of the eigenfunction with the prediction
of the restricted random wave model is quite good.
All moments of the two examples are larger
than those of a Gaussian, corresponding to the larger tails.
Compared to the moments of the restricted random wave model
those of the eigenfunctions tend to be smaller,
in particular for the larger moments. 
This is reasonable, as an actual eigenfunction
is always bounded, which reduces higher moments compared
to the result of eq.~\eqref{eq:moments-via-sigma-q}.

We furthermore have tested our basic assumption 
\eqref{eq:value-distrib-q-restricted-random-wave}, that the local value 
distribution of a sufficiently high lying eigenfunction is Gaussian with 
a variance given by the local classical probability density associated 
with $D$, more directly. To this end we have computed the value distribution 
of the eigenfunction in fig.~\ref{fig:example2} for three small 
regions on which $\var(\BFq)$ is almost invariant, and we therefore expect 
a Gaussian. The results are shown in fig.~\ref{fig:local-fluctuations}, 
and good agreement with the prediction 
\eqref{eq:value-distrib-q-restricted-random-wave} is found. Since many
fewer 
wavelengths are contained in these small domains than in 
$\Omega$ the statistics is of course not as good as for the full system, 
but the results give strong support for a local Gaussian behavior. 
The variances for the two domains $A$ and $C$ coincide with 
the expected classical one $\var(\BFq)$. But for domain $B$ the 
observed variance is larger than $\var(\BFq)=0$. This corresponds to the 
widening of the delta peak in fig.~\ref{fig:example2}, and is due to the 
fact that the eigenfunction cannot become exactly 
zero on some open set at finite energies, but instead fluctuates around zero.

\FloatBarrier

%%%%%%%%%%%%%%%%%%%%%%%%%%%%%%%%%%%%%%%%%%%%%%%%%%%%%%%%%%%%%
\section{Summary}
%%%%%%%%%%%%%%%%%%%%%%%%%%%%%%%%%%%%%%%%%%%%%%%%%%%%%%%%%%%%%

In this work we have extended the random wave model for eigenfunctions 
from the case of chaotic systems to the case of irregular eigenfunctions 
in systems with mixed phase space. Our main result 
is one particular prediction of this model, namely  the 
amplitude distribution \eqref{eq:amp-distrib-mixed}
of irregular eigenfunctions. Numerical tests 
have been performed for two high lying eigenfunctions of the 
\limacon billiard with $\varepsilon=0.3$, and impressive 
agreement, even in the tails of the distribution,
with the theoretical prediction was found. 

The physical picture underlying our analysis is that the local 
hyperbolicity in the irregular part of the phase space forces the 
eigenfunctions localizing on this part of phase space 
to behave locally like a Gaussian random function with a variance 
given by the classical probability density in position space 
defined by the uniform measure on the irregular component. 
Taking the mean over all these local Gaussians with varying 
variance gives our result for the global amplitude distribution. 
We have tested this intuitive picture by computing local amplitude 
distributions.
The agreement of these with the Gaussian prediction 
is very good, giving 
further strong  support to the picture of local Gaussian fluctuations 
with variance determined by the underlying classical system.  
A natural further question relates to the correlations 
of such eigenfunctions between
different points in position space; this topic is addressed in 
\cite{BaeSch2001a:p}.

Although we have restricted our study to Euclidean billiards, the 
general picture of local Gaussian fluctuations is of course not 
limited to these special type of systems. 
We therefore expect our results to be 
valid for irregular eigenfunctions in arbitrary systems 
(e.g.\ systems with potential), with 
$\var(\BFq)$ defined as the projection of the ergodic measure on 
the irregular component to the position space.

\vspace{1cm}

\noindent{\bf Acknowledgements}

\vspace{0.25cm}

\noindent
We would like to thank Francesco Mezzadri for pointing out a shortening
of the original derivation and Professor Jonathan Keating for useful comments
on the \mbox{manuscript}.
A.B.\ acknowledges support by the 
Deutsche Forschungsgemeinschaft under contract No. DFG-Ba 1973/1-1.
R.S.\ acknowledges support by the 
Deutsche Forschungsgemeinschaft under contract No. DFG-Ste 241/7-3.

%%%%%%%%%%%%%%%%%%%%%%%%%%%%%%%%%%%%%%%%%%%%%%%%%%%%%%%%%%%%%%%%%%%%%%%%%%%%%

\end{document}